\documentclass[12pt,a4paper]{article}

\usepackage{SGpreprint}
\usepackage{url,lastpage}

\nologohead{Thomas Fent, Patrick Groeber, Frank Schweitzer\\
 	Coexistence of Social Norms based on
In- and Out-group Interactions \\
\textsl{ACS - Advances in Complex Systems}, vol. \textbf{10}, no. 2
(2007), pp. 271-286}

\usepackage{epsfig}
\usepackage{amsmath}
\usepackage[sort&compress]{natbib}

\begin{document}

\begin{center}

   \textbf{\Large Coexistence of Social Norms based on\\
In- and Out-group Interactions}\\[5mm]
   \textbf{\large Thomas Fent\footnote{present address: 
Vienna Institute of Demography, Austrian Academy of Sciences, 
A-1040 Vienna}, Patrick Groeber and Frank Schweitzer}\\
        Chair of Systems Design\\
        ETH Zurich, Kreuzplatz 5, 8032 Zurich, Switzerland \\
         \texttt{fschweitzer@ethz.ch}
\end{center}

\begin{abstract}

  The question how social norms can emerge from microscopic interactions
  between individuals is a key problem in social sciences to explain
  collective behavior. In this paper we propose an agent-based model to
  show that randomly distributed social behavior by way of local
  interaction converges to a state with a multimodal distribution of
  behavior. This can be interpreted as a coexistence of different social
  norms, a result that goes beyond previous investigations. The model is
  discrete in time and space, behavior is characterized in a continuous state
  space.  The adaptation of social behavior by each agent is based on
  attractive and repulsive forces caused by friendly and adversary
  relations among agents.  The model is analyzed both analytically and by
  means of spatio-temporal computer simulations. It provides conditions
  under which we find convergence towards a single norm, coexistence of
  two opposing norms, and coexistence of a multitude of norms. For the
  latter case, we also show the evolution of the spatio-temporal
  distribution of behavior.

\end{abstract}


\section{Introduction}
\label{introduction}

How individual behavior is determined or at least influenced by social
norms is one of the classic questions of social theory.  Here we consider
a norm as a rule guiding individual decisions concerning rituals,
beliefs, traditions, and routines.  Populations of individuals or
sometimes even companies or nations often exhibit a remarkable degree of
coordinated behavior helping to prevent or govern conflicts.  When this
coordination is enforced without the help of a central authority, the
coordinated behavior and the arising regulation of conflict may be due to
the existence of norms.  What distinguishes a norm from other cultural
products like values or habits is the fact that adherence to a social
norm is enforced by sanctions. As \citet{axelrod} states it: ``A norm
exists in a given social setting to the extent that individuals usually
act in a certain way and are often punished when seen not to be acting in
this way.''  Therefore, the existence of a norm is not a matter of yes or
no but a matter of degree.  In turn, how often a certain action is taken
or how often an actor is punished for not taking that action determines
the growth or decay of a norm.

A social norm can persist although the initial rational origin changes or
even vanishes over time.  Actions that were originally performed because
they were necessary to survive under certain environmental conditions may
continue to persist as a social norm although the current circumstances
do not require them anymore.  Thus, a norm may or may not have a rational
foundation.  Norms are sometimes unwritten and unspoken rules that become
apparent only when they are violated.  Nevertheless, in some societies
norms are clearly defined rules.

Adherence to norms is enforced by sanctions which may be formal or
informal.  For instance, in politics, civil rights and civil liberties
are not only supported by the power of the formal legal system but as
well by informal norms determining what is acceptable \citep{axelrod}.
Then again, violation of a norm may be punished on a purely informal
level in such a way like stigmatizing or ignoring the violator.  Typical
sanction mechanisms used in real life are ostracism, physical
retaliation, refusal of social approval, gossip, etc.
\citep{diekmann2003}.  In the course of development of a society it may
happen that norms become internalized such that violation of norms is
psychologically painful for the deviator even when the sanction mechanism
is not active anymore \citep{scott1971}.  If a norm is internalized by
every member of a society the norm remains stable even without performing
any sanction.  Another possibility of enforcing a social norm is given by
considering one special type of behavior to be the ``normal'' situation,
e.g. in a certain society a leading position can only be assigned to a
man, people above a certain age are assumed to be married and the like.
Consequently, the existence of a social network is a prerequisite for
successful implementation of social norms.

Although norms determine individual behavior they must be negotiated on
the macro level \citep{haferkamp}. Different subgroups of a society
possess different abilities to transfer their local guidelines to other
groups. Basically, the more resourceful groups may allocate resources to
less resourceful groups who will support the institutionalization of a
certain norm. In the sequel both groups internalize the norm.  The
resourceful also have the power to sanction deviation which stabilizes
the norm and further increases the power of the resourceful. However, not
all groups within a society will adopt a certain norm. Individuals may
consider themselves associated with an inclusive group (in-group) but
also have the desire to dissociate from certain other groups of
individuals, the out-groups. This interplay of association and
dissociation on one hand strengthens solidarity within in-groups, but on
the other hand allows for coexistence of contradicting norms within a
society. Consequently, in one and the same situation, the expectations
regarding a certain desired behavior differs among members of different
groups \citep{saam}.

\citet{axelrod} investigates the emergence and stability of behavioral
norms within an $n$--person game.  The players can choose to defect and
receive a payoff for defection.  In the next step, those players who
catch the defector out have the opportunity to impose a punishment but
have to bear the enforcement costs.  However, if this punishment is
costly, a norm to cooperate will not necessarily be established.  Each
strategy has two dimensions determining the players propensity to defect
and the probability to punish deviant behavior.  The actors are endowed
with limited rationality and apply an evolutionary approach to choose
their strategy.  They observe each other and the more successful
strategies are more likely being imitated.  Numerical simulations reveal
that this setup basically does not support the emergence and stability of
a norm suggesting cooperative behavior.  Since no one has any incentive
to punish a defection, the question arises how a norm can ever get
established.  Therefore, Axelrod employs a metanorm ensuring that agents
must punish those whom they detected not punishing observed deviant
behavior.  With this extension a norm against defection is established
and stable once it is established.

\citet{diekmann2003} showed that rational actors in a one--shot situation
are able to enforce social norms with sanctions even when the punishment
is costly.  Many papers address the presence of such social norms.  For
instance \citet{palivos2001} observes the effects of a presence of
family--size norms which indicate that an agent's fertility behavior
depends on prices and income as well as on the fertility rate of the
cohorts.  \citet{lindbeck1999} investigate the interplay between social
norms and economic incentives.  They consider a continuum of individuals
facing the decision to work or to live off public transfers.  Those
individuals who refuse to work receive a transfer but also suffer from
embarrassment due to social stigma.  This disutility increases as the
share of people refusing to work decreases.  Thus, the strength of the
social norm that the source of an individuals means of subsistence should
be the individuals own work is determined endogenously within the
modelling framework.  The model investigated by Lindbeck allows for two
possible outcomes: a low--tax society determined by a majority of
taxpayers or a high--tax society carried by a majority of transfer
recipients.

\citet{cole1992} analyzed a multi--generation model in which parents can
improve their children's matching prospects by increasing savings.  If
all families do that the offsprings' advantage vanishes since their
parents activities offset each other.  Nevertheless, the system is not in
an equilibrium if all families abandon this effort since in such a
situation it would be advantageous for any single family to deviate.
\citet{cole1992} showed that there exist equilibria where over--saving
takes place as well as equilibria where it is suppressed.  In an extended
version of this model \citet{cole} include a {\em wealth--is--status
  social norm}, which means that a woman receiving multiple proposals
accepts the one from the wealthiest candidate, and an {\em aristocratic
  social norm} where a man's status is inherited. While the former social
norm leads to over--saving and deadweight losses the latter allows to
suppress over--saving within families belonging to the upper class.

Another promising field of application of social norms is the
investigation of life course events.  Certainly, the timing and
sequencing of major events of an individuals life course, such as the
first sexual relationship, union formation, leaving parental home,
marriage, and first birth is determined by decisions which are in
principle taken by the individual.  Nevertheless, the individuals'
environment has an influence on these decisions.  This influence may take
place through normative guidelines providing some rules of thumb
generated by the society as a whole but also through imitation of the
behavior of the individuals who are closely connected --- the relevant
others. Neglecting these influence mechanisms, that is, not to behave
according to the rules may incur some costs for the individual such as
the exclusion from a group or the loss of reputation. Therefore, the
normative rules guiding the timing of major life course events are
enforced by formal and informal sanctions.  This qualifies the guidelines
to serve as perceived social norms shaping individuals' lives.
\citet{billari2004} did an empirical in-depth analysis of perceived
norms regarding lower and upper limits on sexual debut and marriage.

\citet{billari2002} introduce an agent--based one--sex
non--overlapping--generations model to understand the dynamics of the
intergenerational transmission of age--at--marriage norms. The social
norms at first influence the agents mate search decisions. In case of a
successful search resulting in a marriage the norms of the partners are
transmitted to their offsprings by means of a certain combiner creating a
new norm for the child based on the parents' norms.  \citet{aparicio}
investigate whether these results also hold in a more complex setup where
heterogeneity with respect to age and sex is explicitly taken into
account.  Moreover, they also include the timing of union formation and
fertility into the model.  To create a more realistic model of the
evolution of age norms the characteristics of the agents are extended and
the social norms are split into two sex-specific norms.

The age--at--marriage norms serve as guidelines for individuals to make
decisions about the right point in time to get married. Normative
guidelines generally are a decision guidance whenever an individual has
to decide about something important. Thus certain actions are influenced
by social norms or social rules that state how individuals ought to
behave in certain circumstances.

The individual being in the situation of taking a decision at the micro
level is guided by social norms imposed at the macro level. Moreover, the
set of all micro level decisions within a certain society generates the
macro level behavior of the system which may either strengthen the
existing social norms or weaken them if there is a collective trend to
deviate. Thus, the long run development of social norms is the result of
collective dynamics within a social network. The society is a system
containing a large number of individuals interacting through their social
networks to serve their own needs.  \citet{granovetter1973,
  granovetter1983, granovetter1985} provides a theory of embeddedness
suggesting that all economic action accomplished either by individuals or
by organizations is enabled, constrained, and shaped by social ties among
individuals.  The number of connections may vary among individuals but we
may assume that there is no completely unconnected individual (except the
man in the moon) and no one is connected to all others.  The impact of
different types of connectivity, i.e. the influence of the network
structure under consideration has been extensively studied (see for
instance \citet{barabasi1999}, \citet{collins}, \citet{rahmandad}, and
\citet{watts1998nature}).

\citet{ehrlich} emphasize that human beings are not only the result of
biological evolution but also of a process of cultural evolution.  In
opposite to genes which can only pass unidirectionally from one
generation to the next, norms, ideas, conventions, and customs can pass
between individuals distant from each other and even from the children to
their parents.  Ehrlich and Levin postulate that a clear understanding of
the interactions between cultural changes and individual actions is
crucial to the success of efforts to influence cultural evolution.
Cooperation in human societies relies essentially on social norms even in
modern societies, where cooperation substantially hinges on the legal
enforcement of rules.  A theory of social norms should help to explain
how norms emerge, how they are maintained, and how one norm replaces
another.  Moreover, we do not only want to discuss individual behavior in
the presence of norms but also how norms change over time.

The remainder of this paper is organized as follows. Section \ref{model}
explains the simulation model we developed to investigate the evolution
of norms within a population of artificial agents, followed by a detailed
discussion of several model aspects in section \ref{ingroup}. In section
\ref{results} we present and discuss the results obtained in various runs
of numerical simulations and in section \ref{summary} we summarize and
interpret these results.

\section{The model}

\label{model}

We consider an artificial population featuring $N$ agents.  Each agent $i
\in \{1,\ldots,N\}$ is linked to other agents from his in-group $I(i)$ and
his out-group $O(i)$. That means within the social network of agent $i$
we distiguish between two different subgroups: members of an agent's
in-group can be seen as his friends, whereas the members of the
out-group are regarded as persons with whom he has adversary relations
(enemies).  The number of agents in $I(i)$ is given by $k_{i} := |I(i)|$
and the size of $O(i)$ is $l_{i} := |O(i)|$.  The behavior of agent $i$
at time $t$ is denoted by $x_{i}^{t} \in [0,1]$ and the current behavior
of all agents within an in-group determines the group's social norm.

We further assume that $I(i)$ and $O(i)$ are always disjoint and that the
in-/out-group relation is symmetric, e.g. $j\in I(i)\Leftrightarrow i\in
I(j)$ and $j\in O(i)\Leftrightarrow i\in O(j)$.  So for example an agent
can neither belong to another agent's in- \textit{and} out-group nor not
be included in another agent's in-group (or out-group respectively) if he
conversely belongs to the latter agent's in-group (out-group).

If agents $i$ and $j$ belong to the same in-group but deviate from each
other, they receive (and impose) a punishment which increases with the
difference $x_{i}^{t} - x_{j}^{t}$ between their social behavior. In
order to ensure a symmetric punishment and to ease the model's analytical
tractability, we simply choose the square of that difference, $(x_{i}^{t}
- x_{j}^{t})^{2}$.  Consequently, agent $i$ receives a disutility for
deviating from his in-group members' behavior which is proportional to
$\sum_{j \in I(i)} (x_{i}^{t} - x_{j}^{t})^2$.  Moreover, the agents are
reluctant to change their own behavior, which is characterized by a
disutility proportional to $(x_{i}^{t+1} - x_{i}^{t})^2$.  Finally, each
group of the population has the desire to express its own identity.
Therefore, agents obtain a positive utility by differing from the
out-group proportional to $\sum_{j \in O(i)} (x_{i}^{t} - x_{j}^{t})^2$.
We assume that agent $i$ can only observe the current behavior within the
population but does not have the ability to anticipate future movements
of other agents. So to determine the utility maximizing behavior
$x_i^{t+1}$ at time $t+1$, he uses the previous behavior $x_j^t$ of his
in-/out-group members (including his own behavior $x_i^t$).

Introducing the parameter $\alpha \in [0,1]$
to adjust the weight of his utilities and disutilities caused by his own
opinion on the one hand and his in- and out-group members on the other
hand, the utility function which agent $i$ wants to maximize becomes
\begin{eqnarray}
   U(x_{i}^{t+1}) & = &  - \alpha (x_i^{t+1} - x_i^t)^2  \nonumber \\
& & + (1-\alpha)\left[-\sum_{j \in I(i)} (x_{i}^{t+1} - x_{j}^{t})^{2}
    + \sum_{j \in O(i)} (x_{i}^{t+1} - x_{j}^{t})^{2} \right].
\label{utility}
\end{eqnarray}
Note that hence the higher $\alpha$ is, the more an agent will be
punished for deviating from his behavior at the previous timestep.  On
the other hand the outer influence from in- and outgroup will increase
with growing group-sizes $k_i$ and $l_i$.  

Assuming that an agent cannot foresee the impact of his own decision on
the other agents' behavior, the partial derivatives of (\ref{utility})
become
\begin{eqnarray}
  \frac{\partial U(x_{i}^{t+1})}{\partial x_{i}^{t+1}} & = &
  - 2 \alpha (x_i^{t+1}-x_i^t)  \nonumber \\ & & 
+ 2(1-\alpha)\left[-\sum_{j \in I(i)} (x_{i}^{t+1} - x_{j}^{t}) +
    \sum_{j \in O(i)} (x_{i}^{t+1} - x_{j}^{t})\right],
  \label{utility'}
  \\
  \frac{\partial^{2} U(x_{i}^{t+1})}{(\partial x_{i}^{t+1})^{2}} & = &
  -2 \left[\alpha + (1 - \alpha)(k_{i} - l_{i}) \right] .
\end{eqnarray}
If the utility function is strictly convex (i.e. its second derivative is
positive), the optimal $x_i^{t+1}$ must be either zero or one. But this
means every agent's choice will always be one of these extreme values
which seems inappropriate for further consideration as the spectrum of
behavior would be reduced to only two possible values.  For this reason
we assume
\begin{equation}
\label{concave}
   \alpha + (1-\alpha)(k_i - l_i) > 0.
\end{equation}
It is sufficient (but of course not necessary) for this condition that
every agent's in-group consists of at least as many members as his
out-group does. Otherwise we must choose $\alpha$ close enough to one.

With assumption (\ref{concave}) the utility maximizing $x_{max}$ becomes
\begin{equation}
\label{update}
   x_{max} = \frac{
      \alpha x_{i}^{t} + (1 - \alpha) (\sum_{j \in I(i)} x_{j}^{t} - \sum_{j \in O(i)} x_{j}^{t})}
      {\alpha + (1 - \alpha)(k_{i} - l_{i})
   }.
\end{equation}
As $x_i^{t+1}\in[0,1]$ is required, we set
\begin{equation}
  x_i^{t+1} =
  \begin{cases}
    x_{max} \quad & \text{for } x_{max}\in [0,1] \\
    0 & \text{for } x_{max} < 0\\
    1 & \text{for } x_{max} > 1
  \end{cases}
\end{equation}
to choose the optimal value within this interval and by this define the
dynamics in a way that always $x_{i}^{t+1}\in [0,1]$. If $x_{max} < 0$,
we choose the left border of that interval, and if $x_{max} > 1$ the
right one.

\section{The isolated in-group mechanism}
\label{ingroup}

\subsection{Costs of sanctions}

Now we will have a look at the costs of being punished and at the costs
of imposing a punishment.  Recall that in the simple version of the
\citet{axelrod} simulation model the agents are reluctant to impose a
punishment since there is no economic incentive to punish and it even
incurs costs.  However, as \citet{fehr2004} pointed out, sanctions are
the decisive factor for norm enforcement.  Anyhow, in the real world
individuals are willing to impose a punishment even if this is
disadvantageous in economic terms as long as the costs of imposing a
sanction are not very high.  In an experimental setup deployed by
\citet{fehr2004a} a third party observes test persons in a prisoners'
dilemma and has the option to punish players for defecting.  Although
disadvantageous from a purely profit--maximizing point of view third
parties are willing to punish defection particularly when the opponent
cooperated.  Thus, the enforcement of norms is largely driven by
nonselfish motives.  These findings may empirically justify Axelrod's
approach to include a metanorm.  In our model, we exercise a similar
approach by just taking it for granted that people are punished and
impose a punishment, respectively, if agents deviate from the behavior of
their in-group (recall equation (\ref{utility})).  Nevertheless, a
social norm will only be enforced by sanctions if the costs of punishing
are much lower than the costs of being punished.  In principle, the
sanction mechanism in our model is totally symmetric as in case of
deviation of two agents they will receive an identical decrease of
utility.  But having a high majority for a certain opinion in a group
offers a slightly different interpretation: Let us assume a fully
connected group of individuals without any further links to outer agents.
Moreover, since we are only looking at the sanction but not at the desire
to deviate from the out-group, we assume them to be empty so that all
links with the group represent in-group connections.  We further assume
having an agent deviating from the homogeneous rest of the group, i.e. we
have the two types of behavior $x_{1}$ and $x_{j}=x$ for $j \in \{2,
\ldots , n\}$. Because of its dominance within the group the latter
opinion could be considered as this group's norm.

If agent $1$ refuses to converge toward the other agents (i.e. $x_{1}^{t}
= x_{1} \forall t$) and the other agents refuse to converge as well, he
receives a disutility
\begin{equation}
   U(x_{1}) = - (1 - \alpha) (n - 1) (x_{1} - x)^{2}
\end{equation}
from being punished while the other agents have to bear the costs
\begin{equation}
   U(x_{j}) = - (1 - \alpha) (x_{1} - x)^{2}
\end{equation}
for imposing the punishment. Therefore, the disadvantage of being
punished is $(n-1)$ times higher than the enforcement costs.

Considering the isolated out-group mechanism (empty in-groups)
in an analogous situation, every agent would receive a positive utility
$\tilde{U}(x_i)=-U(x_i), 1\leq i\leq n$ for deviating from his out-group.
So the change in utility caused by deviation can be interpreted as a reward
instead of a punishment here.

If the agents maximize their utility in a scenario with empty out-groups,
we can conclude from equation
(\ref{update}) that their behavior in the next time step becomes
\begin{eqnarray}
   x_{1}^{t+1} & = & \frac{
      \alpha x_{1}^{t} + (1 - \alpha)(n - 1) x^{t}
   }{\alpha + (1 - \alpha)(n - 1)} \\
   x_{j}^{t+1} & = & \frac{
      (1 - \alpha) x_{1}^{t} + [\alpha + (1 - \alpha)(n - 2)] x^{t}
   }{\alpha + (1 - \alpha)(n - 1)}, \quad 2\leq j\leq n.
\end{eqnarray}
From that it follows that the deviator (agent $1$) makes a $(n-1)$ times
bigger movement than the other group members. For $\alpha = 1/2$ the
agents converge to a common behavior already after one iteration.

\subsection{Status within a group}

In real populations the status of an individual determines his power and
influence and also his propensity to adhere to social norms.  Individuals
with a higher status gain more from community membership which also
increases the threat of ostracism. If an individual gains little or
nothing from community membership the threat of ostracism is of little
importance \citep{cole}.  In this simulation the number of links an agent
possesses represents his status within the population.  As we can see from
the utility function or directly from the dynamics in (\ref{update}), the
possible punishment and therefore an agent's change in behavior increases
with the size of his in-group. Hence, the number of connections determines
the influence of the individual on the behavior of the population but
also the number of people who can punish an agent for deviating from
their own behavior.  Consequently, agents with a higher status are more
interested in corresponding to their relevant others than those with a
low status.

\subsection{Connections to other models}

Note that in case every agent's out-group is empty (e.g. $k_i=0$), we
have $x_{max}\in[0,1]$ and the dynamics can be written as
\begin{equation}
x^{t+1} = A x^t
\end{equation}
and therefore
\begin{equation}
x^{t+1} = A^{t+1}x^0
\end{equation}
with $x^t=(x_1^t,\dotsc,x_N^t)^T$, $x^0$ representing the initial
behavior and a $n\times n$-matrix $A=(a_{ij})$ defined by
\begin{equation}
a_{ij} =
\begin{cases}
\frac{\alpha}{\alpha + (1-\alpha)k_i} \quad & \text{for } i=j \\
\frac{1-\alpha}{\alpha + (1-\alpha)k_i} \quad & \text{for } i\neq j,\quad i\in I(j) \\ 
0 \quad & \text{for } i\neq j,\quad i\notin I(j)
\end{cases}.
\end{equation}

One can easily verify that $A$ is row-stochasticm, i.e. in any row, its components
sum up to one. Thus omitting theout-group mechanism we get a
special case of the model introduced by \citet{degroot1974}.

So if we only consider the punishment of deviation from the in-group, an
agent's behavior is a weighted mean of his own and his in-group members'
behavior one time step before. The dynamics thereby only depend on $A$
and its powers.  If the powers of $A$ converge to a matrix with identical
rows, all agents will finally adopt the same global social norm.
Sufficient and necessary conditions for this can be taken from
\citet{hegselmann}.  Hence the asymptotic behavior of the agents
qualitatively only depends on the in-group structure and its representing
graph or adjacency matrix.  A change of the initial vector $x^0$ or the
parameter $\alpha$ within the interval $(0,1)$ would only affect the
value of the limit but not the qualitative asymptotic behavior.

In absence of out-group interactions, our model is also connected in both its
formulation and some limiting results to the model of \citet{deffuant00a} whereas
the agents interact pairwise in the latter case.
Furthermore, \citet{jager04} also provide attracting and repulsive forces
in continuous opinion dynamics as an extension of \citet{deffuant00a}.
There the agents attract each other if their current distance in behavior is below
a threshold while repulsion occurs if this distance is above a second threshold.
In our model, not the agents' current state but the in-/out-group structure
determines whether attraction or repulsion takes effect on them.

\section{Simulation Results}
\label{results}

\begin{figure}[htbp!]
  \begin{center}
    \includegraphics[width=0.5\textwidth]{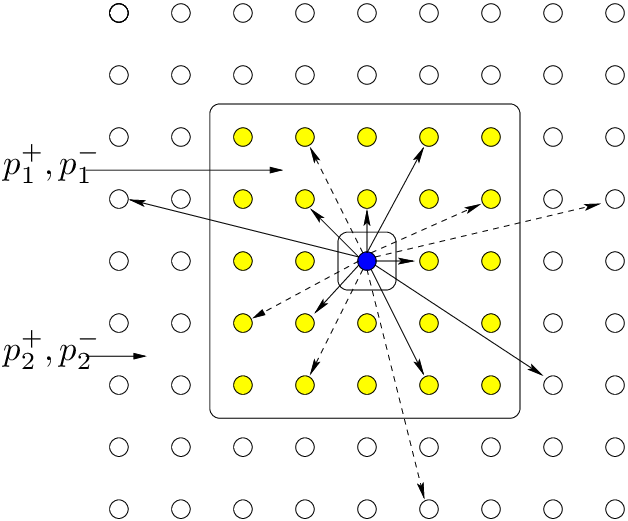}
  \end{center}
  \caption{Two agents are in each other's in-group (out-group) with a
    probability $p_1^+$ ($p_1^-$) if they are neighbors with respect to
    the Moore-Neighborhood of size $2$.  Otherwise the probability is
    $p_2^+$ ($p_2^-$). Solid lines illustrate the central agent's in-group relations while
    his out-group relations are represented by dashed lines.
    Note that $p_1^+,p_1^- > p_2^+,p_2^-$, i.e. there are more relations to neighbors than
    to agents outside the neighborhood.}
  \label{Neighborhood}
\end{figure}

Evidently, our model's dynamics heavily depend on the in-/out-group
structure of the agents which can be arbitrary complex. In our simulation
we generate random in- and out-groups considering the spatial relation
between the agents which is defined by an additional network, in our
setting a two dimensional lattice with periodic boundary conditions. The
spatial effect is brought into play by assuming that for an agent it is
more likely to have (positive or negative) relations to other agents in a
certain neighborhood compared to far distant agents. Thus with $N(i)$
denoting the neighborhood for agent $i$, we define 
\begin{eqnarray}
\nonumber p_1^+ = & P(j\in I(i) &| j\in N(i)), \\
\nonumber p_1^- = & P(j\in O(i) &| j\in N(i)), \\
\nonumber p_2^+ = & P(j\in I(i) &| j\notin N(i)), \\
\label{in-out-P}
p_2^- = & P(j\in O(i) &| j\notin N(i))
\end{eqnarray}
as the probabilities for an agent $j\neq i$ to be in the in- or out-group
of agent $i$. We further assume $p_1^+, p_1^- > p_2^+, p_2^-$ which means
that it is more likely for an agent to be in another one's in- or
out-group if he is in that agent's neighborhood (see Figure
\ref{Neighborhood}). Based on these probabilities, we construct a random
in-/out-group structure for every agent.

\begin{figure}[htbp!]
  \begin{minipage}[c]{0.5\textwidth}
    a)
    \begin{center}
    \includegraphics[scale=0.3]{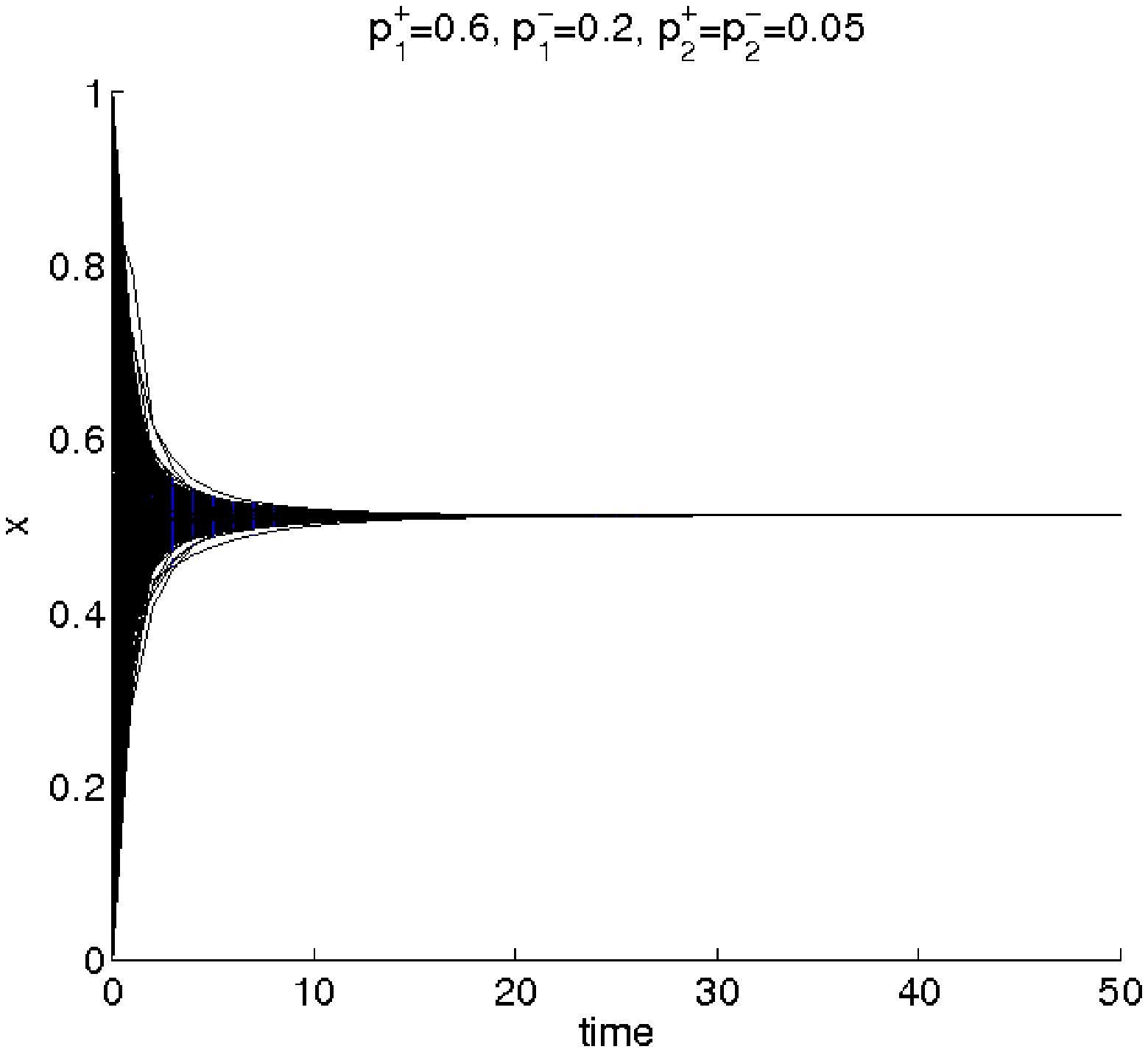}
    \end{center}
  \end{minipage}%
  \begin{minipage}[c]{0.5\textwidth}
    b)
    \begin{center}
   \includegraphics[scale=0.3]{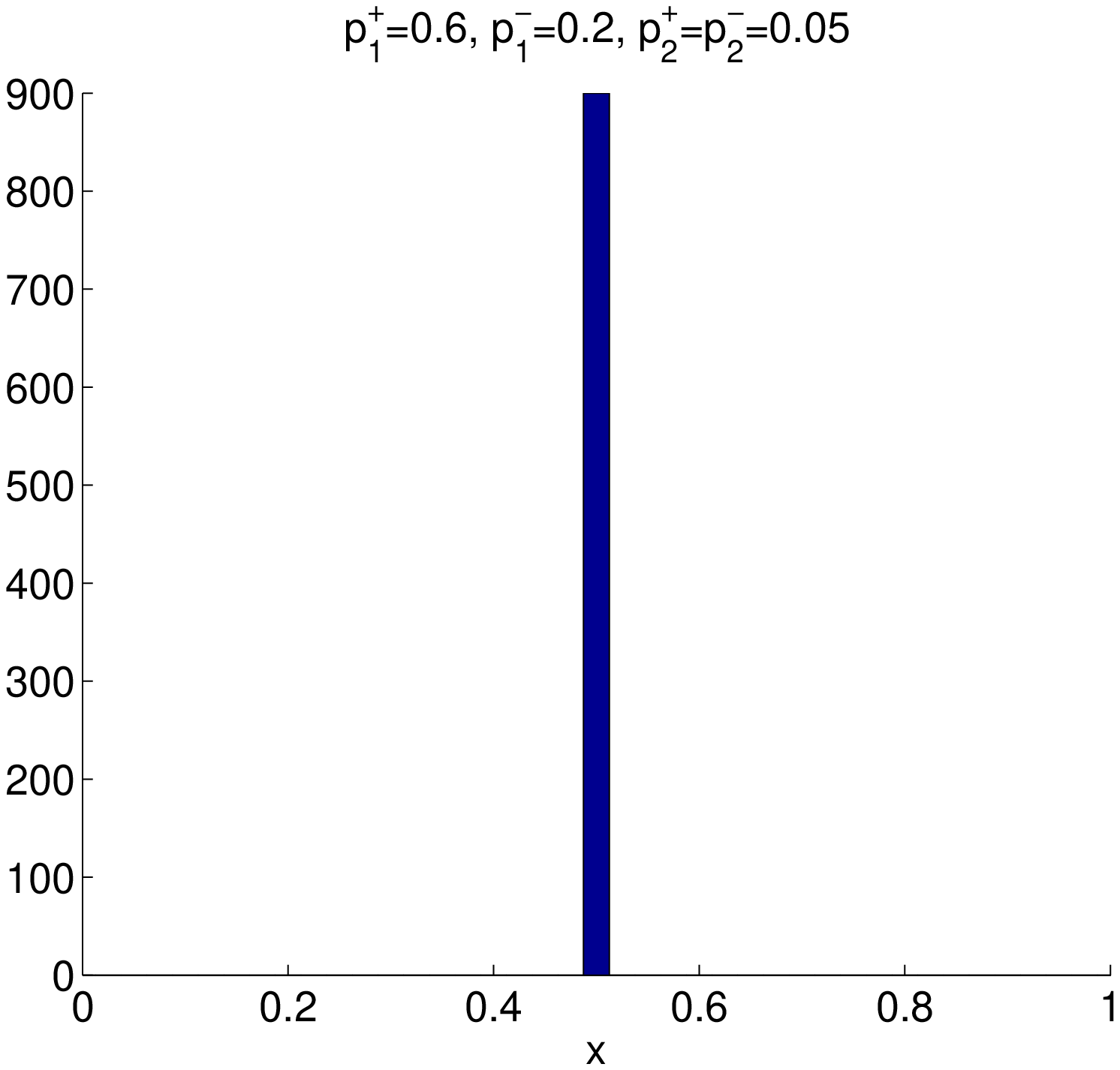}
    \end{center}
  \end{minipage}%
  \ \\
  \begin{minipage}[c]{0.5\textwidth}
    c)
    \begin{center}
      \includegraphics[scale=0.3]{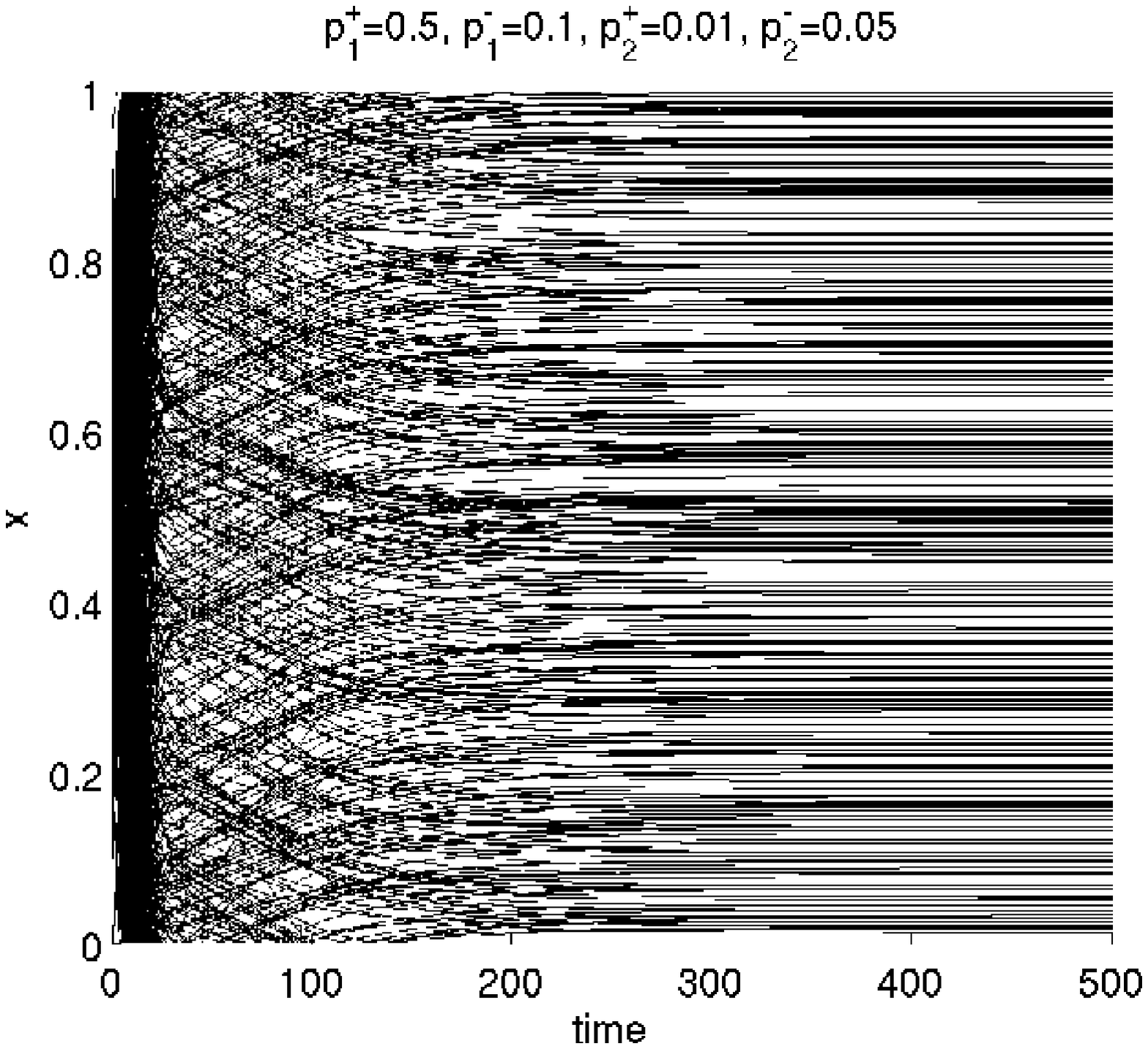}
    \end{center}
  \end{minipage}%
  \begin{minipage}[c]{0.5\textwidth}
    d)
    \begin{center}
      \includegraphics[scale=0.3]{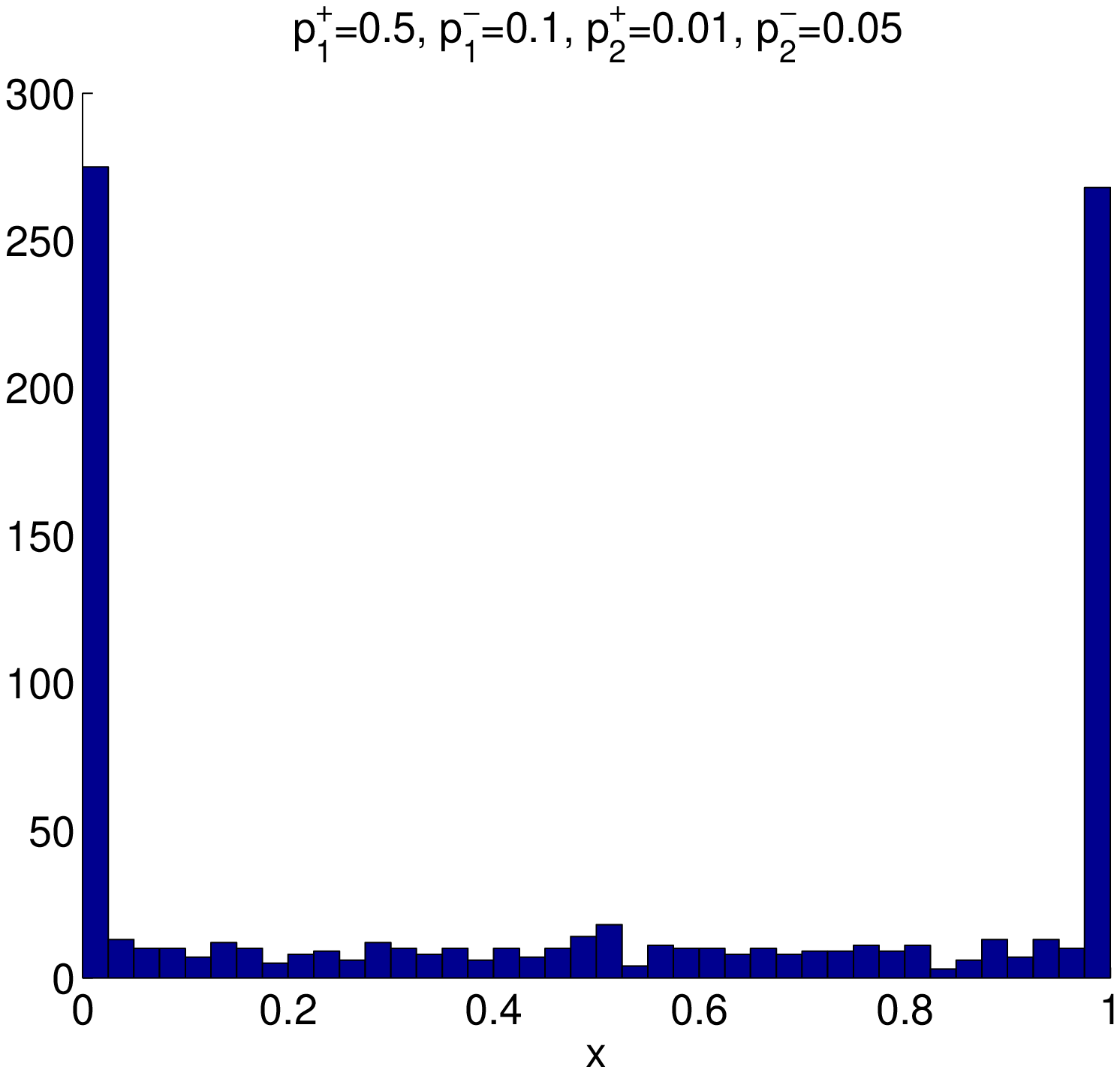}
    \end{center}
  \end{minipage}%
  \ \\
  \begin{minipage}[c]{0.5\textwidth}
    e)
    \begin{center}
      \includegraphics[scale=0.3]{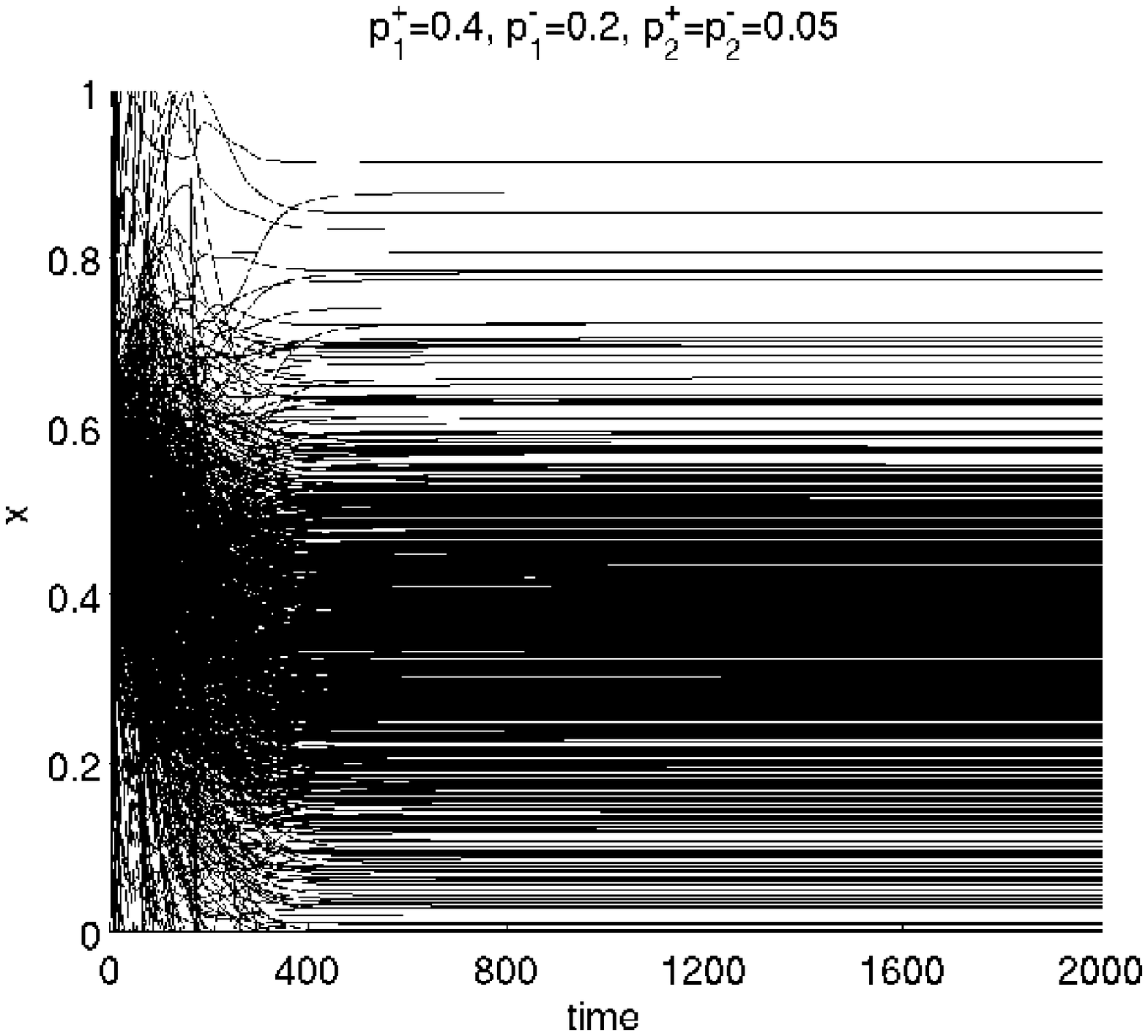}
    \end{center}
  \end{minipage}%
  \begin{minipage}[c]{0.5\textwidth}
    f)
    \begin{center}
      \includegraphics[scale=0.3]{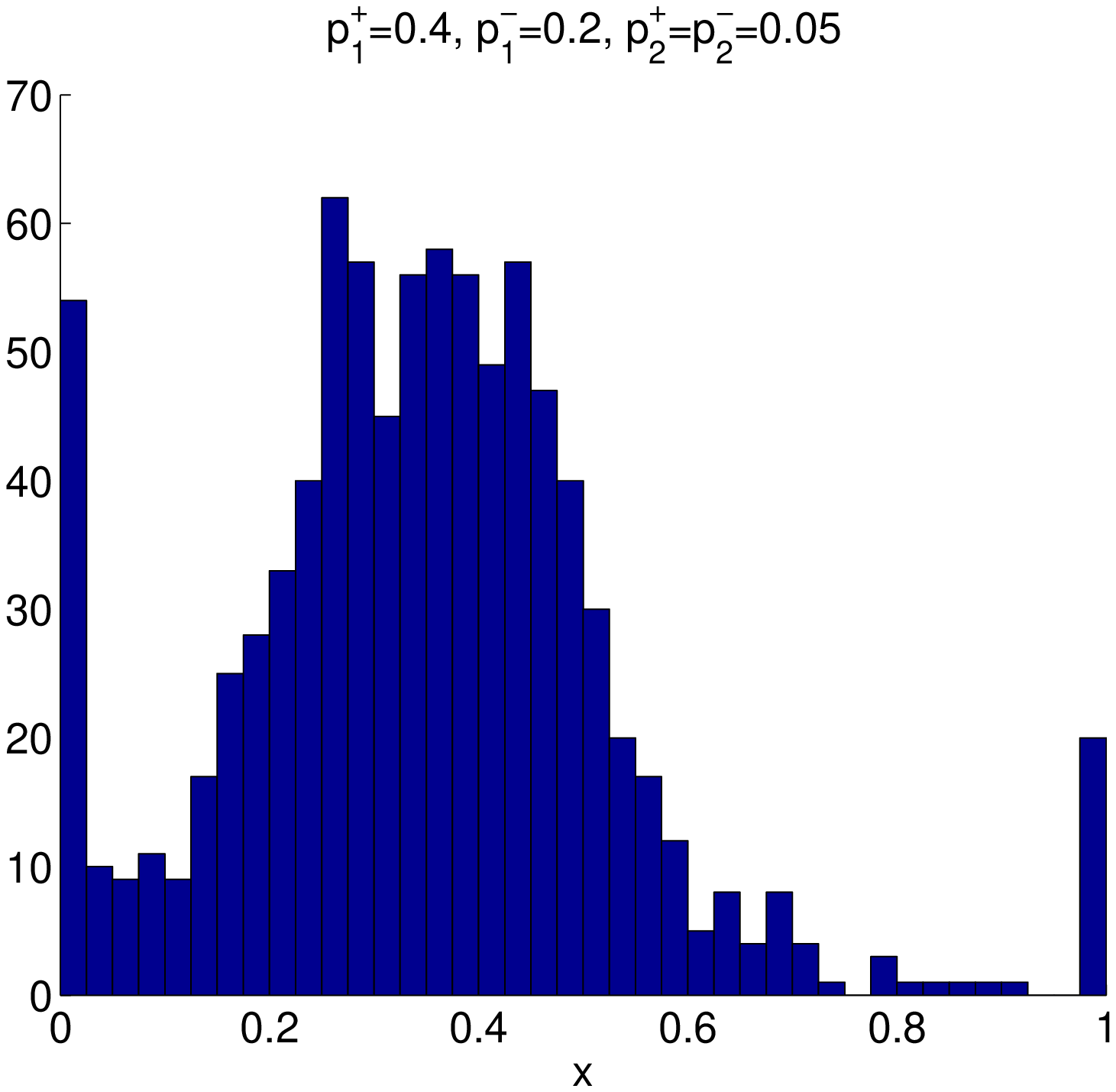}
    \end{center}
  \end{minipage}%
  \caption{Trajectories and distribution of the agent's behavior $x_{i}$
    after the last step of simulation: (a,b): 50, (c,d): 500, (e,f): 2000
    time steps.}
\label{Trajectories_Histograms}
\end{figure}

\begin{figure}[htbp!]
  \begin{minipage}[c]{0.5\textwidth}
    a)
    \begin{center}
      \includegraphics[scale=0.35]{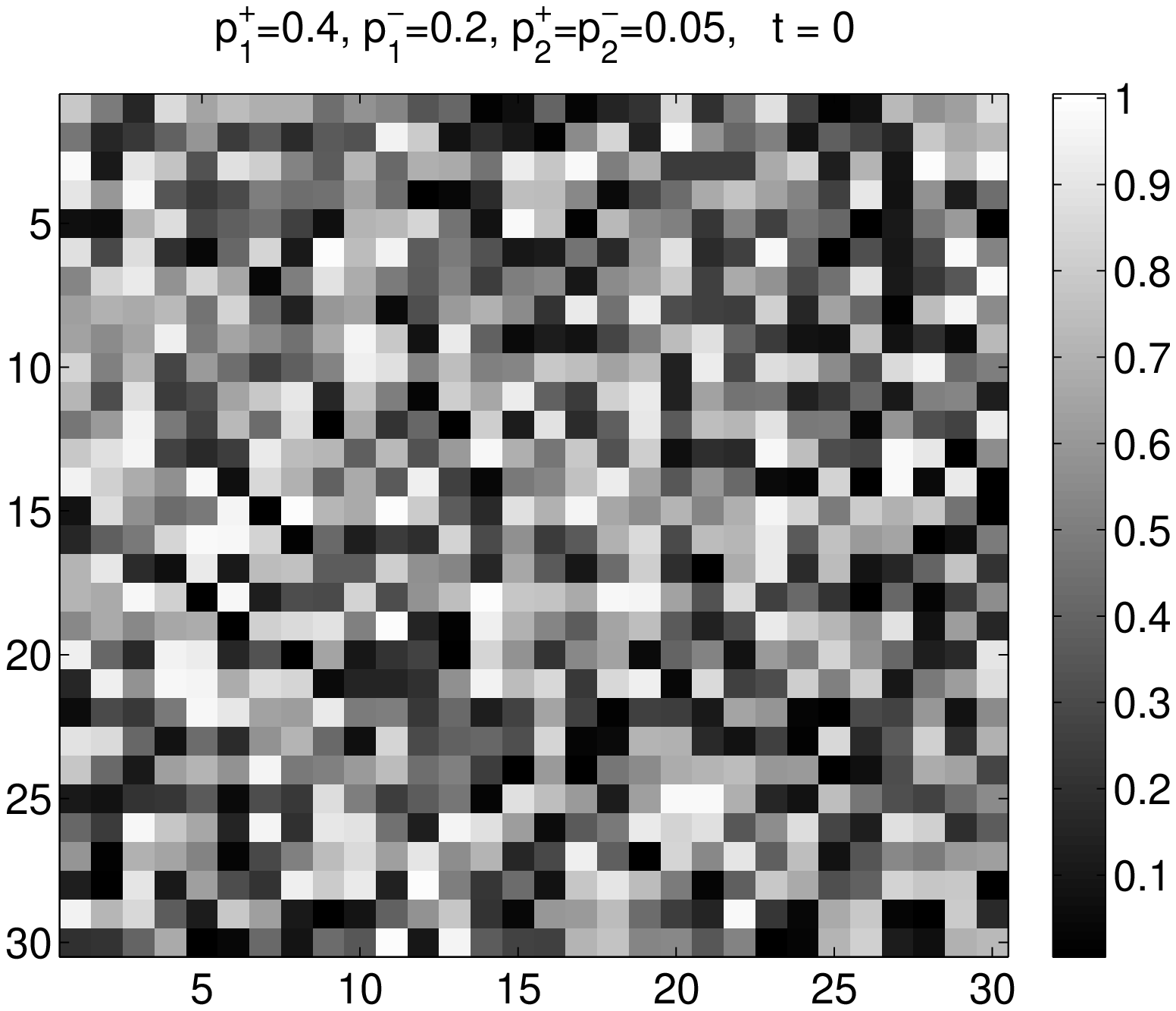}
    \end{center}
  \end{minipage}%
  \begin{minipage}[c]{0.5\textwidth}
    b)
    \begin{center}
      \includegraphics[scale=0.35]{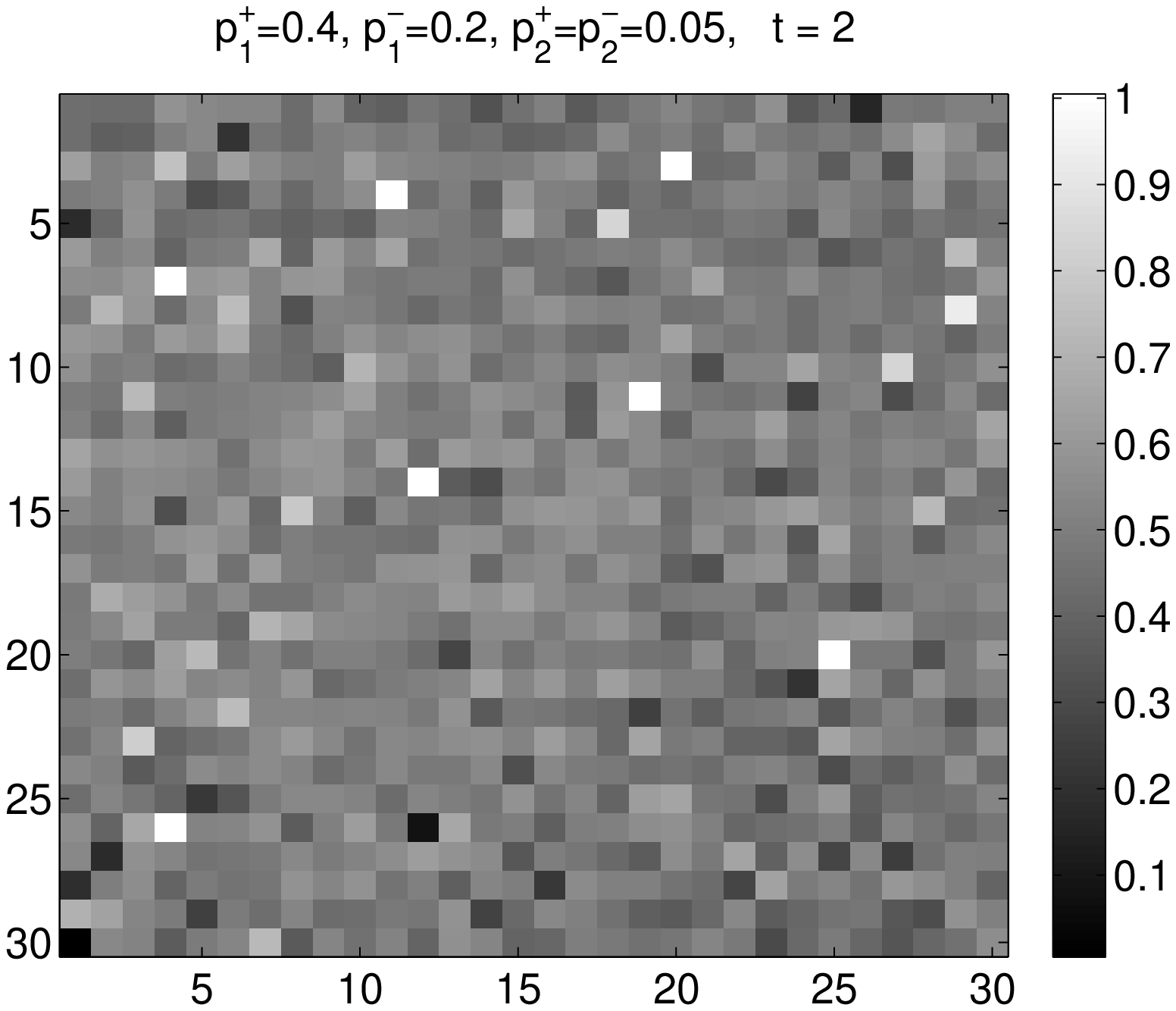}
    \end{center}
  \end{minipage}%
  \ \\
  \begin{minipage}[c]{0.5\textwidth}
    c)
    \begin{center}
     \includegraphics[scale=0.35]{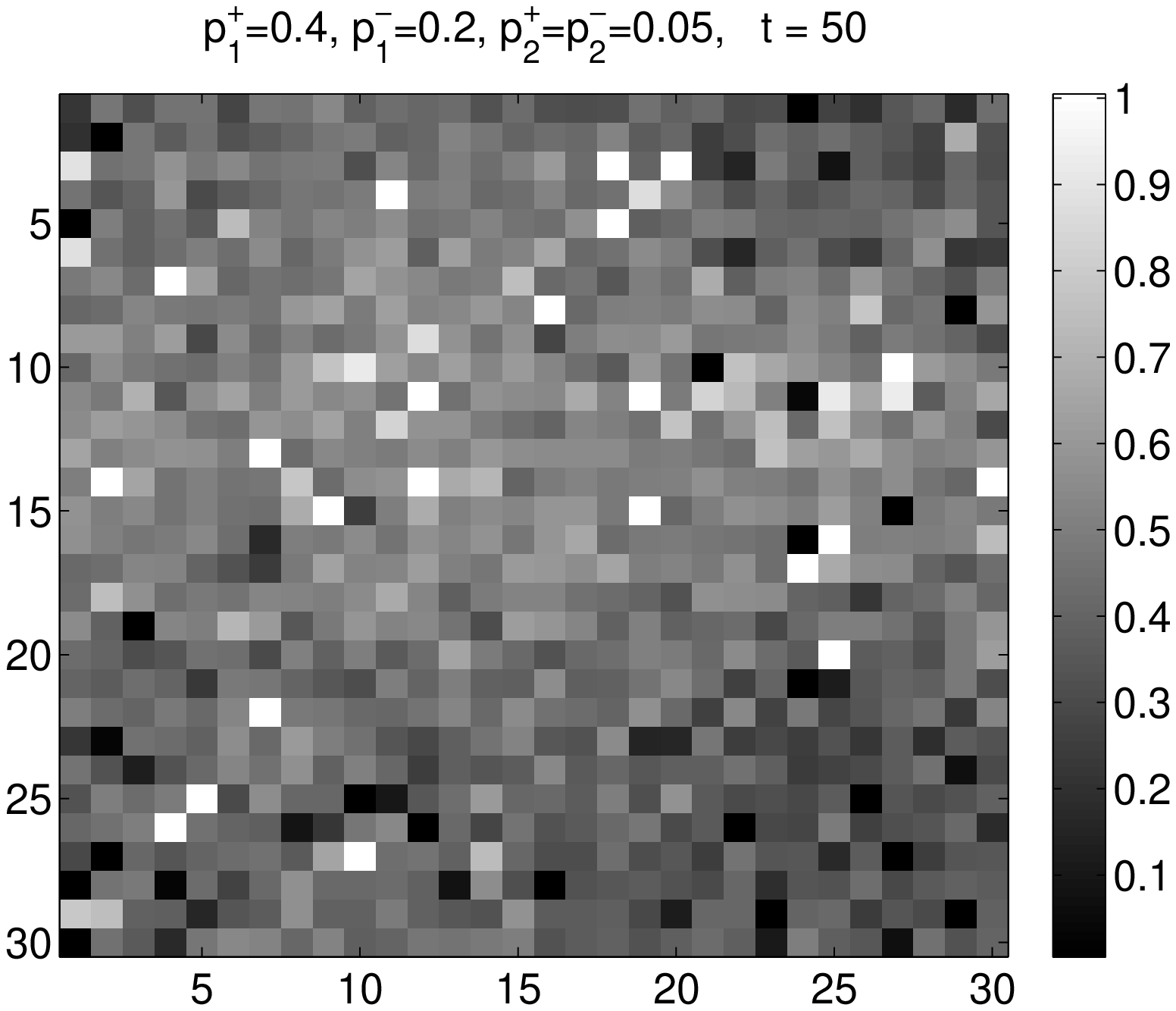}
    \end{center}
  \end{minipage}%
  \begin{minipage}[c]{0.5\textwidth}
 d)
    \begin{center}
  \includegraphics[scale=0.35]{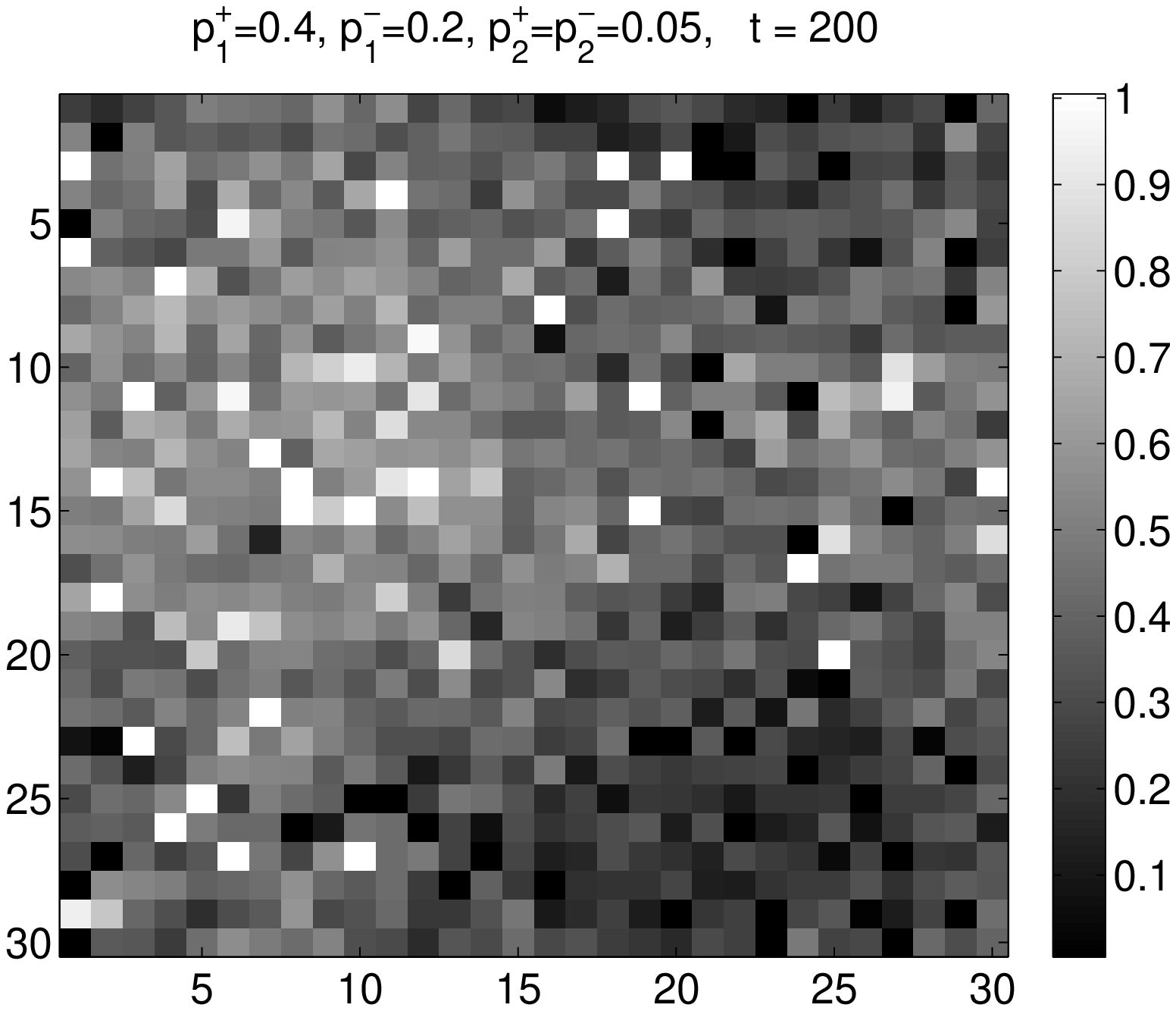}
    \end{center}
  \end{minipage}%
  \ \\
  \begin{minipage}[c]{0.5\textwidth}
  e)
    \begin{center}
     \includegraphics[scale=0.35]{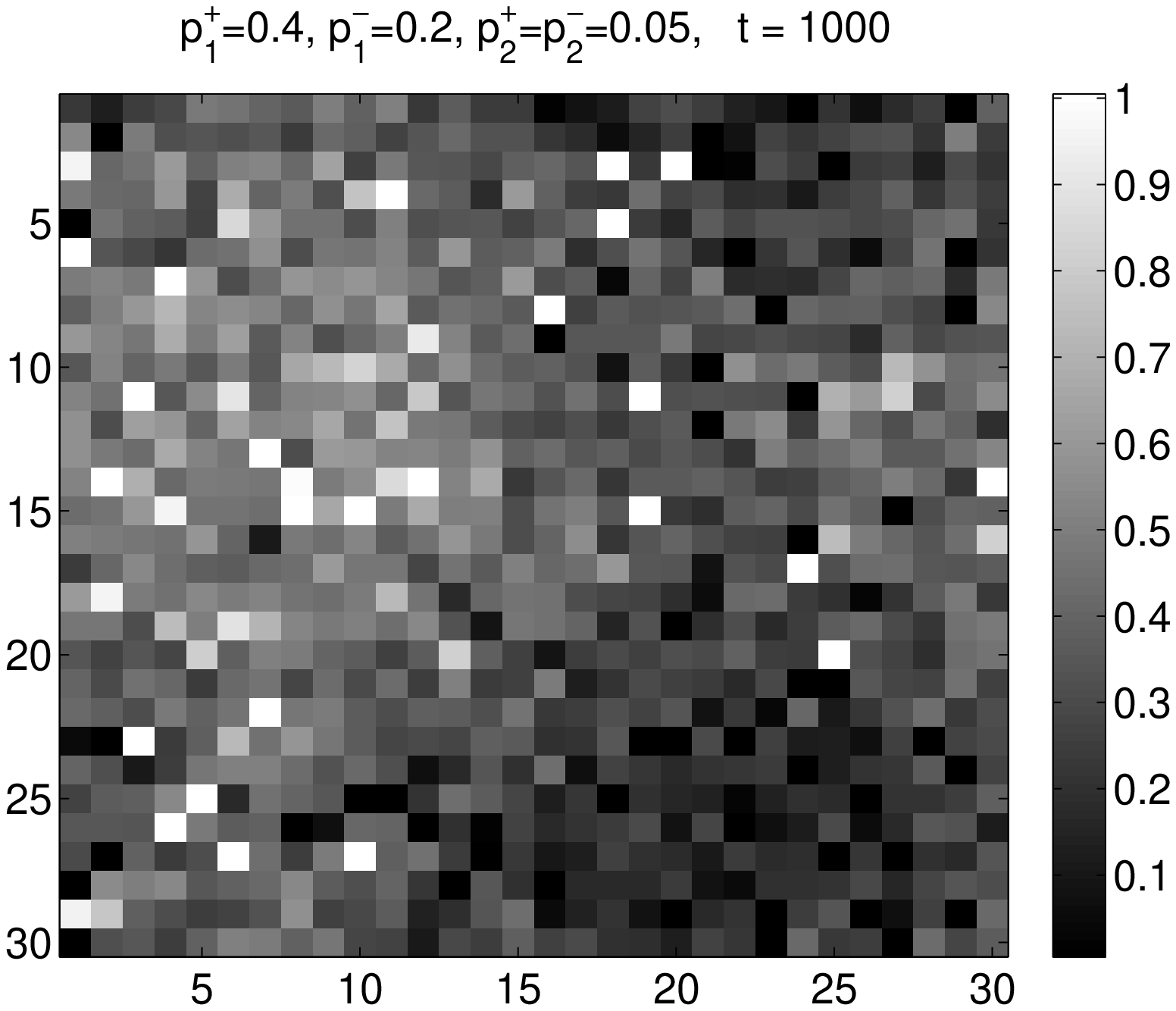}
    \end{center}
  \end{minipage}%
  \begin{minipage}[c]{0.5\textwidth}
  f)
    \begin{center}
    \includegraphics[scale=0.35]{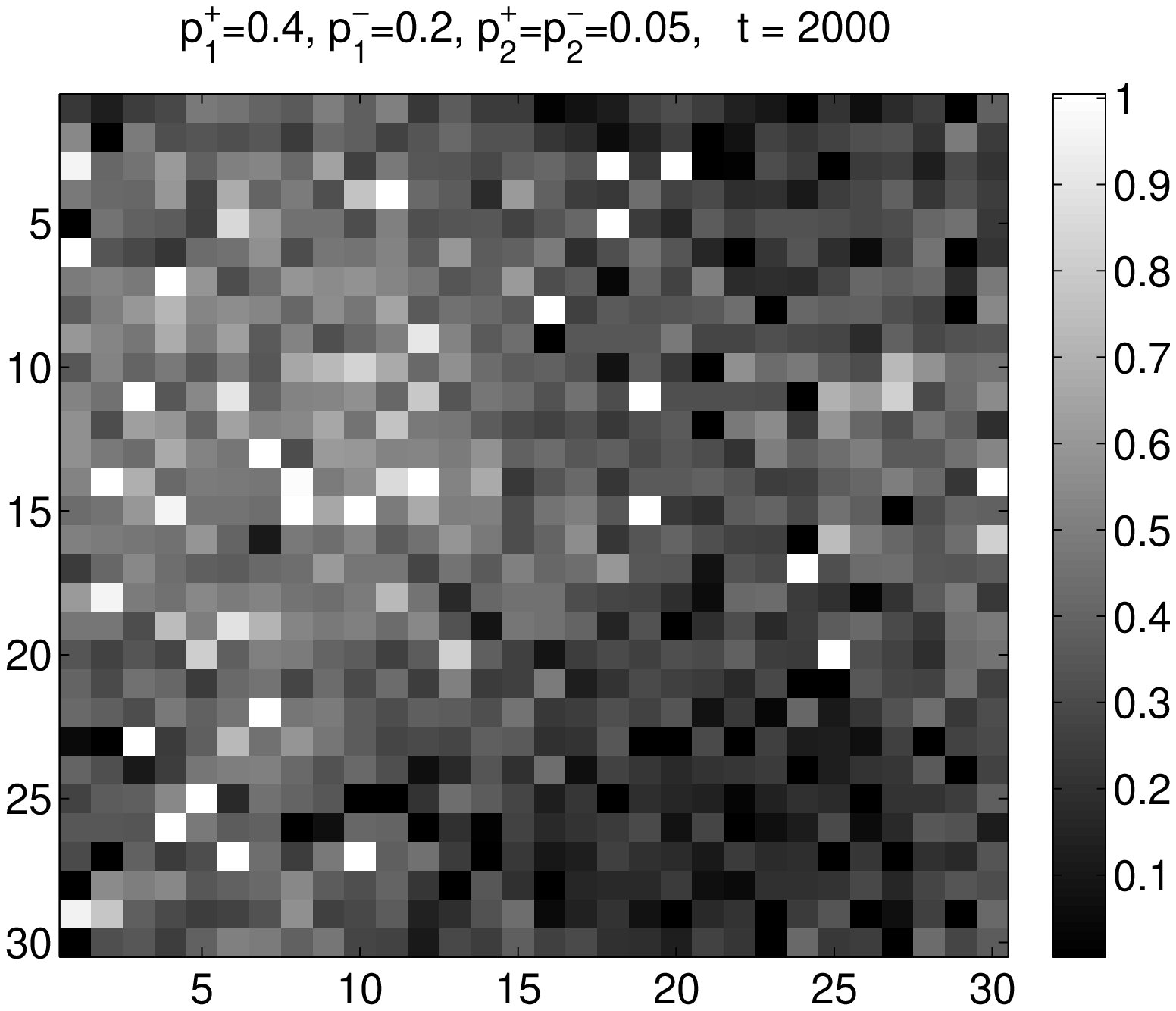}
    \end{center}
  \end{minipage}%
  \caption{Spatio-temporal evolution of the agent's behavior $x_{i}$,
    which is coded in grey scale. The probabilities for the in- and
    outgroup relations within and outside the neighborhood are chosen as
    $p_1^+=0.4$, $p_1^-=0.2$, $p_2^+=p_2^-=0.05$, which allows for the
    spatial coexistence of a multitude of different norms. The
    corresponding distribution is shown in Fig.
    \ref{Trajectories_Histograms}.f.  From this histogram it is clear
    that there are at least three dominating peaks, which can be
    interpreted as different social norms.}
\label{movies}
\end{figure}

In our simulations we always consider $N=900$ agents, each with a
Moore-neighborhood of size $13 \times 13$, with the agent placed in the
center.  For $\alpha=0.9$, Figure \ref{Trajectories_Histograms} shows the
agents' trajectories and the distribution of their behavior after the
last timestep of simulation for in-/out-group realizations for different
values of the probabilities from equation (\ref{in-out-P}). The first two
examples show more simple cases: in Fig.  \ref{Trajectories_Histograms}.a
and Fig.  \ref{Trajectories_Histograms}.b we observe that the agents very
quickly converge to a consensus very close to 0.5, so here the spectrum's
center is the norm all agents finally conform to.  This case always
appears if the size of the agents' in-group is large compared to their
outgroup-size.

Fig.  \ref{Trajectories_Histograms}.c and Fig.
\ref{Trajectories_Histograms}.d show a setting where we find a wider
spectrum of behavior.  Here, every agent has about 5 times more friends
than enemies in his neighborhood while outside of this the ratio is vice
versa (with lower absolute numbers).  So it is more likely for an agent
to accord with his neighbors which also makes a more likely emergence of
local clusters plausible. The simulations show that in this situation, we
find about one third of the agents stabilizing at each end of the
spectrum while the last third is almost equally distributed over $[0,1]$.

The most interesting situation is shown in Fig.
\ref{Trajectories_Histograms}.e and Fig.
\ref{Trajectories_Histograms}.f: the setting is similar to Figs.
\ref{Trajectories_Histograms}.a-b, only $p_1^+$ has been reduced from 0.6
to 0.4.  This reduction of each agent's in-group in his neighborhood
reduces the attracting force between the agents sufficiently to avoid a
global consensus in behavior. Instead of this we find about three peaks
in the distribution: one at the empirical mean of approximately 0.37 and
two at the end of the interval whereas the peak at zero is higher than at
one.

For this situation, Fig. \ref{movies} depicts the spatial evolution of
the agents' behavior over time. Already after two iterations, the initial
random distribution has been evened out to a level close to 0.5 for
almost all agents - this also applies to most other observed settings. So
we have a short timescale where attraction between the agents is
dominant.  On a second, larger timescale we observe a differentiation of
the agents' behavior: After 50 steps we find that agents in the upper
region prefer values greater than 0.5 whereas at the opposite side values
lower than 0.5 are preferred. Furthermore, the empirical mean of the
agents' behavior decreased clearly under the initial value of 0.5. In the
next picture we find the the upper left corner dominated by agents with
values higher than 0.5 while the remaining area the majoritarian behavior
is clearly lower than 0.5 and the overall empirical mean is already lower
than 0.4. As we see in Fig. \ref{movies}.e and Fig. \ref{movies}.f, this
situation remains stable for the rest of the simulation, so neither the
contracting nor the dispersing forces prevail by driving the agents to
total consensus or polarization respectively. An animated computer
simulation (in color) of the whole spatio-temporal evolution shown can be
found at \texttt{http://www.sg.ethz.ch/research}.

Varying the parameter $\alpha$ (without violating equation
(\ref{concave})) to change the weight for an agent's own behavior and
that of his in- and out-group respectively did not change the results
qualitatively and only affected the system's time to reach its stationary
state - a larger $\alpha$ increases an agent's weight on his own behavior
and extends this time.

\section{Conclusions}
\label{summary}

In this paper, we have studied the emergence of social norms to allow for
a better understanding of the self--organized dynamics of social behavior
in human societies.  Our model considers several influences explicitely:
persistence, i.e.  the individuals' reluctance to alter their behavior,
solidarity, the desire to be associated with a certain group (the
in-group), and the desire to differ from some individuals belonging to
the out-group.  These three components have been incorporated in an
agents' utility function to be maximized.  While some literature on
social norms suggest that norm enforcement is driven by nonselfish
motives \citep[e.g.][]{fehr2004a} we consider profit maximizing agents
but explicitly define a disutility obtained from deviations within the
in-group.  Thus, instead of inserting a metanorm like \citet{axelrod}, the
agents in our model are just assumed to experience a disutility from
deviation or bear the costs of imposing a sanction, respectively.

The agents' in- and out-group-structure -- one of the key features of our
model -- is chosen randomly with the restriction that an agent is more
likely linked to neighboring agents than to those outside his
neighborhood. Hence an agent's interaction is not restricted to this
neighborhood, but its influence on him is higher compared to the rest
of the population. Depending on these probabilities we could observe
different effects. The greater the in-group size is compared to the
out-group size, the more the attracting forces between the agents
dominate and lead to global consensus at the center of the spectrum of
behavior. If we increase the out-group size to a certain level, the
attraction between agents is still dominant at the beginning whereas on a
larger timescale we find a differentiation of the agents' behavior
leading to stable clusters with different social norms. Further
increasing of the out-group size results in a polarization of the agent's
behavior with a majority equally distributed over the two extreme values
zero and one.  

We analyzed the model analytically and by means of computer simulations
whereas the simulation is not systematic but restricted to interesting examples.
As a major finding, our model is appropriate to explain the emergence and
the stable spatio-temporal coexistence of different social norms
prevalent in certain subgroups of the society. Further the final distribution of
behavior is smoother compared to the opinion dynamics model of
\citet{jager04} which also provides coexistence
in a spatio-temporal setting caused by attractive and repulsive forces
between the agents. We want to emphasize that
the topology of the social network, in particular the in- and out-group
structure, is crucial for the development of the agents' social behavior.
For the model under consideration we can imagine situations which may not
converge into a (quasi)stationary distribution, i.e. agents change their
social behavior constantly over time and the social norm adjusts
instantaneously. While this kind of scenario may also have some
relevance, we argue that a social norm should change only on time scales
larger than the dynamics of the agents' individual behavior (which is
shown as a quasistationary phenomenon in this framework).

A future extension of the model shall include the fact that in reality an
individual's social network is not static, but changes over time.  In our
model, this can be covered by an explicite dynamics of the probabilities
$p_1^+, p_1^-,p_2^+, p_2^-$. More general, one could also consider
feedbacks between the agents' behavior and the in- and out-group
relations or a network topology which changes endogenously.

\section*{Acknowledgements}
The authors wish to thank Nicole J. Saam (Erfurt) and Andreas Flache
(Groningen) for their comments on a first version of this paper.

\bibliographystyle{plainnat}
\bibliography{norms_arxiv}

\end{document}